\documentclass[%
 aip,
 amsmath,amssymb,
 reprint,%
]{revtex4-1}

\usepackage{graphicx}
\usepackage{dcolumn}
\usepackage{bm}
\usepackage{multirow}
\usepackage{calc}
\usepackage{subcaption}
\usepackage{nicefrac}

\usepackage[utf8]{inputenc}
\usepackage[T1]{fontenc}
\usepackage{mathptmx}
\usepackage{etoolbox}

\makeatletter
\def\@email#1#2{%
 \endgroup
 \patchcmd{\titleblock@produce}
  {\frontmatter@RRAPformat}
  {\frontmatter@RRAPformat{\produce@RRAP{*#1\href{mailto:#2}{#2}}}\frontmatter@RRAPformat}
  {}{}
}%

\usepackage{graphicx}
\usepackage{dcolumn}
\usepackage{bm}
\usepackage[hidelinks]{hyperref}
\usepackage{cleveref}

\crefrangeformat{equation}{Equations~#3#1#4--#5#2#6}
\crefmultiformat{equation}{Equations~#2#1#3}{ and~#2#1#3}{, #2#1#3}{ and~#2#1#3}
\crefformat{equation}{Equation~#2#1#3}
\crefrangeformat{figure}{Figures~#3#1#4--#5#2#6}
\crefmultiformat{figure}{Figures~#2#1#3}{ and~#2#1#3}{, #2#1#3}{ and~#2#1#3}
\crefformat{figure}{Figure~#2#1#3}
\crefrangeformat{table}{Tables~#3#1#4--#5#2#6}
\crefmultiformat{table}{Tables~#2#1#3}{ and~#2#1#3}{, #2#1#3}{ and~#2#1#3}
\crefformat{table}{Table~#2#1#3}
\crefname{section}{Section}{Sections}
\crefname{subsection}{Subsection}{Subsections}
\crefname{appendix}{Appendix}{Appendices}
\makeatother

\newlength{\widthoffourzeros}
\newlength{\widthofpercent}
\begin{document}

\preprint{APS/123-QED}

\title{Testing Koopmans spectral functionals on the analytically-solvable Hooke's atom}

\author{Yannick Schubert}
\email{yannick.schubert@uzh.ch}
\affiliation{Department of Physics, Eidgenössische Technische Hochschule Zürich, 8092 Zurich, Switzerland}
\altaffiliation[Now at ]{Department of Chemistry, University of Zurich, 8057 Zurich, Switzerland}
\author{Nicola Marzari}
\affiliation{Theory and Simulations of Materials (THEOS), École polytechnique fédérale de Lausanne, 1015 Lausanne, Switzerland}
\affiliation{National Centre for Computational Design and Discovery of Novel Materials (MARVEL), École polytechnique fédérale de Lausanne, 1015 Lausanne, Switzerland}
\author{Edward Linscott}
\email{edward.linscott@epfl.ch}
\affiliation{Theory and Simulations of Materials (THEOS), École polytechnique fédérale de Lausanne, 1015 Lausanne, Switzerland}

\date{\today}

\begin{abstract}
Koopmans spectral functionals are a class of orbital-density-dependent functionals designed to accurately predict spectroscopic properties. They do so markedly better than their Kohn-Sham density-functional theory counterparts, as demonstrated in earlier works on benchmarks of molecules and bulk systems. This work is a complementary study where --- instead of comparing against real, many-electron systems --- we test Koopmans spectral functionals on Hooke's atom, a toy two-electron system that has analytical solutions for particular strengths of its harmonic confining potential. As these calculations clearly illustrate, Koopmans spectral functionals do an excellent job of describing Hooke's atom across a range of confining potential strengths. This work also provides broader insight into the features and capabilities of Koopmans spectral functionals more generally.
\end{abstract}

\maketitle

\maketitle

\maketitle

\maketitle
\section{Introduction}



In order to assist the design of optical and electronic devices, it is important to be able to predict spectral properties from first principles. However, it is impossible to simulate these systems exactly. Even without the spin degrees of freedom, the corresponding Schrödinger equation is a differential equation in $3N$ dimensional space, where $N$ is the number of electrons. Solving this equation directly for practically every system of interest vastly exceeds today's (and tomorrow's) computational capabilities, so we must necessarily make some approximations.

There are several different approaches one can take here. High-level many-body perturbation theory \cite{Onida2002} (e.g. the GW approximation \cite{Hedin1965}) and wavefunction-based approaches (e.g. coupled cluster \cite{McClain2017} or Quantum Monte Carlo \cite{Foulkes2001}) are accurate but computationally expensive, which limits what systems we can study with these approaches. A more computationally inexpensive approach is density-functional theory (DFT) \cite{Hohenberg1964a,Kohn1965a}. DFT introduces a framework that reformulates the complex many-body particle problem as a set of $N$ independent single-particle problems (described by the Kohn-Sham equations) and therefore tremendously reduces the complexity. However, DFT is a theory founded on the prediction of total energies, rather than spectroscopic properties, and as such its prediction of spectroscopic properties can be unreliable at best.

Recently, Koopmans spectral functionals \cite{Dabo2009b,Dabo2010,Dabo2014,Borghi2014, Borghi2015, Nguyen2015, Nguyen2016,Nguyen2018, colonna_screening_2018, Degennaro2021, Colonna2022, Linscott2023} have been developed as a beyond-DFT extension designed to improve the prediction of quasi-particle related properties. Previous studies have already demonstrated that Koopmans spectral functionals work well for predicting spectral properties of real molecules and solids when compared to the solutions obtained from accurate quantum chemistry methods, and to experimental results \cite{Colonna2019, Degennaro2021}. However, they have never been tested on an analytically solvable system. With an exact solution at hand the features and capabilities of different functionals can be evaluated in much greater detail. 

In this paper we apply Koopmans spectral functionals on one such analytically solvable system: Hooke's atom. Before presenting the results we will first briefly introduce the theory needed to interpret our findings in \cref{sec: Theory}. This includes a revision of some of the main problems of standard Kohn-Sham density-functional theory (KS-DFT), a motivation how Koopmans spectral functionals address these issues and a quick introduction to Hooke's atom. In \cref{sec: comp-methods} we explain the computational methods used in this study. In \cref{sec: results} we present our results from which we draw our conclusions in \cref{sec: conclusion}.

\section{Theory}\label{sec: Theory}
\subsection{Problems with standard Kohn-Sham DFT \label{sec: summary}}
A spectral theory ought to have quasi-particle energies that match the total energy differences corresponding to electron removal $E(N) - E_i(N-1)$ and addition $E_i(N+1) - E(N)$. This is true for the exact Green's function, whose poles are given directly by these total energy differences, but for KS-DFT the KS orbital energies $\{\varepsilon_j^{KS}\}$ are not theoretically related to quasi-particle energies;  they are simply the single-particle energies of some auxiliary system that happens to have the same ground state density as the system of interest. The exception to this is the highest occupied molecular orbital (HOMO) energy in a finite system which correctly corresponds to the negative ionization potential in the framework of exact KS-DFT \cite{Perdew1982a, almbladh_exact_1985}. Nevertheless, the Kohn-Sham potential is the best local and static approximation to true dynamical and non-local electronic self-energy\cite{Casida1995}, and in practice the KS eigenenergies often qualitatively (and sometimes even quantitiatively) compare well to experiment, and thus it is common practice to interpret KS eigenenergies as approximate quasi-particle energies \cite{chong_interpretation_2002}.

On top of this, errors introduced when approximating the exchange-correlation potential can make Kohn-Sham energies poor approximations to quasi-particle energies. One of the main reasons for this is the erroneous curvature of the total energy as a function of the total number of electrons for non-integer occupations of the orbitals \cite{cohen_fractional_2008} instead of the correct piecewise linear behaviour \cite{Perdew1982a}. 
It follows from Janak's theorem\cite{Janak1978a} 
that the incorrect curvature of the total energy directly affects KS eigenvalues which we would otherwise like to interpret as quasi-particle energies. 

\subsection{Koopmans spectral functionals}
\label{sec: kc-functionals}
Koopmans spectral functionals are a class of functionals that restore the correspondence between the Kohn-Sham energies and total energy differences, by imposing the condition that the derivative of the total energy with respect to the occupation of each orbital must be independent of the occupation of the orbital itself \cite{Dabo2010}:
\begin{equation}
    \varepsilon^\mathrm{Koopmans}_i = \left\langle \phi_i \right| \hat H \left| \phi_i \right\rangle = \text{constant with respect to} \ f_i
\end{equation}
We call this the ``generalized piecewise linearity" (GPWL) condition. It is a sufficient but not a necessary condition to fulfil the piecewise linearity condition discussed in the previous section (which pertains to the total number of electrons, as opposed to the occupation of individual orbitals).

This GPWL is imposed on a DFT energy functional $E^\mathrm{DFT}$ (the ``base'' functional, such as the LDA or PBE), via a Koopmans correction term. This correction removes, orbital-by-orbital, the non-linear dependence of the energy $E$ on the orbital occupation $f_i$ and replacing it with a term that is linear in $f_i$:
\begin{equation}
    E^\mathrm{Koopmans}
    = E^{DFT}
    + \sum_i
    \left[
    -\left(
    E^\mathrm{DFT}
    - \left.E^\mathrm{DFT}
    \right|_{f_i = 0}
    \right) + f_i \eta_i
    \right]
    \label{eqn:pi_i}
\end{equation}
where $\left.E^\mathrm{DFT}\right|_{f_i=f}$ is the energy of the ($N-1+f$)-electron system with the occupation of orbital $i$ constrained to be $f$. $\eta_i$ is some yet-to-be-determined constant. The first term counteracts any dependence of $E^\mathrm{DFT}$ on $f_i$ and the second term replaces it with an explicit linear dependence. 

The above construction of the corrective term is difficult to evaluate directly, unless we only account for the explicit dependence of the orbital density on its occupancy. That is to say, if we have an orbital density $\rho_i(\mathbf{r}) = f_i |\phi_i(\mathbf{r})|^2 = f_i n_i(\mathbf{r})$ we assume that $\phi_i(\mathbf{r})$ is independent of $f_i$ and hence $\rho_i(\mathbf{r})$ is linear in $f_i$. In this case
\begin{equation}
    \left.E^\mathrm{DFT}\right|_{f_i=f} = E^\mathrm{DFT}[\rho - \rho_i + f n_i]
\end{equation}
This frozen-orbital picture allows us to evaluate the terms in the correction, but we cannot neglect the effect of orbital relaxation. In order to account for orbital relaxation \emph{post hoc}, the Koopmans corrections are scaled by scalar prefactors $\alpha_i \in [0, 1]$. These screening parameters are not fitting parameters but can be calculated from first principles in order to impose the GPWL condition. This can be achieved either by finite difference calculations \cite{Nguyen2018,Degennaro2021} or via density functional perturbation theory \cite{Colonna2022}.

This brings us to the final form of the Koopmans functional:
\begin{align}
    & E^\mathrm{Koopmans}[\{\rho_i\}] \nonumber \\
    & \quad = E^{DFT}[\rho]
    + \sum_i
    \alpha_i
    \left[
    -\left(
    E^\mathrm{DFT}[\rho]
    - E^\mathrm{DFT}[\rho - \rho_i]
    \right) + f_i \eta_i
    \right]
    \label{eqn: koopmans_energy_functional}
\end{align}
(Note that the total density is related to the orbital densities via $\rho = \sum_i \rho_i$.) This end result is an orbital-density-dependent functional theory (ODDFT), because now the total energy is a functional not of the total density but of the densities of individual orbitals $\{\rho_i\}$. 

The effect of these corrections is that the total energy is linear in each orbital occupancy $f_i$. However, there is one degree of freedom in this construction: we must choose appropriate slopes $\eta_i$ for these terms in \cref{eqn: koopmans_energy_functional}. The most natural choice is to construct $\eta_i$ such that these slopes correspond to the $\Delta$SCF total energy differences as given by the base DFT functional:
\begin{align}
    \eta_i^\mathrm{KI} = & \left.E^\mathrm{DFT}\right|_{f_i = 1} - \left.E^\mathrm{DFT}\right|_{f_i = 0} \nonumber \\
    = & E^\mathrm{DFT}[\rho - \rho_i + n_i] - E^\mathrm{DFT}[\rho - \rho_i]
\end{align}
where the second equality adopted the frozen-orbital approximation. This is called the Koopmans integer (``KI") functional. Alternatively, the so-called ``KIPZ'' functional additionally includes a screened Perdew-Zunger one-body self-interaction correction, where the screened Hartree-plus-xc energy is removed orbital-by-orbital \cite{Perdew1981a}. This additionally ensures that the KIPZ functional is self-interaction free for one-electron systems \cite{Borghi2014}.

Given this construction, the energy $\varepsilon_i$ of orbital $i$ is related to the difference in total energy of the $N$-electron system ($E^\mathrm{DFT}(N)$) and the total energy of the system where we have added / removed one electron ($E_i^\mathrm{DFT}(N\pm 1)$) calculated at the level of the DFT base functional. These energy differences in turn can be seen as approximations of the true energy $\tilde{\varepsilon}_i$ of the quasi-particle $i$:
\begin{align}
\tilde{\varepsilon}_i&\equiv \Delta E_i \equiv E(N)-E_i(N-1)\nonumber\\
& \approx E^\mathrm{DFT}(N)-E^\mathrm{DFT}_i(N-1)\notag\\
&=\frac{E^\mathrm{DFT}(N)-E^\mathrm{DFT}_i(N-1)}{N-(N-1)}\notag\\
&\overset{\mathrm{GPWL}}{=} \frac{d E^\mathrm{Koopmans}}{d  f_i}\notag\\
&= \varepsilon_i^\mathrm{Koopmans}\notag
\end{align}
for occupied states and
\begin{align}
\tilde{\varepsilon}_i&\equiv \Delta E_i \equiv E_i(N+1)-E(N)\nonumber\\
& \approx E_i^\mathrm{DFT}(N+1)-E^\mathrm{DFT}(N)\notag\\
&=\frac{E_i^\mathrm{DFT}(N+1)-E^\mathrm{DFT}(N)}{N+1-N}\notag\\
&\overset{\mathrm{GPWL}}= \frac{d E^\mathrm{Koopmans}}{d  f_i}\notag\\
&= \varepsilon_i^\mathrm{Koopmans}\notag
\end{align}
for empty states. Therefore, GPWL allows the KS energies to be interpreted as quasi-particle energies.

In summary, Koopmans spectral functionals are orbital-density-dependent functionals that impose a generalized piecewise linearity condition. By constructing corrective terms parameterized in terms of $\Delta$SCF calculations, the resulting Kohn-Sham eigenvalues match the corresponding $\Delta$SCF total energy differences, which in the framework of DFT are formally meaningful and practically reliable.

There is an importance caveat when it comes to the reliability of $\Delta$SCF results. It is well-known that for large molecules and bulk systems, $\Delta$SCF total energy differences cease to be reliable and instead asymptotically approach the corresponding Kohn-Sham eigenvalue. This stems from the fact that the Kohn-Sham eigenstates are very delocalized and removing a single electron from such an orbital starts to resemble the derivative of the total energy with respect to the orbital occupancy. Koopmans spectral functionals overcome this issue by using a basis of localized orbitals. By constructing the Koopmans corrections via the total energy differences that result from the removal of a localized electronic density, the Koopmans corrections --- even though they are based on $\Delta$SCF energy differences --- remain accurate for large and even infinite systems. For more details, refer to Ref.~\citenum{Nguyen2018}. 

\subsection{Hooke's Atom} 
The goal of this paper is to gain insight into the features and capabilities of Koopmans spectral functionals by testing them on Hooke's atom. This is a system where two electrons are confined in an external harmonic potential \cite{Kestner1962}. The Hamiltonian of this system is
\begin{align}
\mathcal{H} = -\frac{1}{2}\left(\nabla_1^2+\nabla_2^2\right)+\frac{1}{2}\omega^2\left(r_1^2+r_2^2\right)+\frac{1}{r_{12}}
\label{eqn: hamiltonian}
\end{align}
where $r_{12} = |\mathbf{r}_1 - \mathbf{r}_2|$. The strength of the confining harmonic potential is given by the parameter $\omega$. Unlike most two-electron systems, it is possible to solve this system analytically\cite{Taut1993} --- although this is only true for particular values of $\omega$. The largest finite value with an analytical solution is $\omega=\nicefrac{1}{2}$, where the exact ground state wavefunction is given (modulo a normalization constant) by
\begin{align}
\Psi\left(\mathbf{r}_1,\mathbf{r}_2;\omega=\nicefrac{1}{2}\right)&= \left(1+\frac{r_{12}}{2}\right)e^{-\left(r_1^2+r_2^2\right)/4} \label{eqn: gs_half}
\end{align}
and it has an energy of 
\begin{align}
    E\left(\omega = \nicefrac{1}{2}\right) = 2 \text{ Hartree} \label{eq: tot_half}
\end{align}
and a ground-state density of
\begin{align}
\rho(\textbf{r})&=2\int d\textbf{r}_2\left|\Psi(\textbf{r},\textbf{r}_2)\right|^2\nonumber \\
&=2 e^{-r^2/2}\left\{\left(\frac{\pi}{2}\right)^{1/2}\left[\frac{7}{4}+\frac{1}{4}r^2\right.\right.\nonumber\\
& \qquad \left.\left.+\left(r+\frac{1}{r}\right)\text{erf}\left(2^{-1/2}r\right)\right]+e^{-r^2/2}\right\} \label{eqn: gs_density_half}
\end{align} 

The next largest value with an analytical solution is $\omega = \nicefrac{1}{10}$. Here, the total energy is 
\begin{align}
    E\left(\omega = \nicefrac{1}{10}\right) = 0.5 \text{ Hartree} \label{eq: tot_tenth}
\end{align}
 and
\begin{align}
\Psi\left(\textbf{r}_1,\textbf{r}_2;\omega=\nicefrac{1}{10}\right)= & \left(1+\frac{r_{12}}{2}
+\frac{r_{12}^2}{20}
\right) e^{-\left(r_1^2+r_2^2\right)/20} \label{eqn: gs_tenth}
\end{align}

Each of these analytical solutions corresponds to instances where a power series expansion of the radial wave function terminates after a finite number of terms. For the general derivation of these analytical solutions see Ref.~\onlinecite{Taut1993}.

In addition to these solutions, one can consider the physics of Hooke's atom in various limiting cases. Rescaling lengths by $\omega^{-1/2}$ and energies by $\omega$, \cref{eqn: hamiltonian} becomes
\begin{align}
\mathcal{H} = -\frac{1}{2}\left(\nabla_1^2+\nabla_2^2\right)+\frac{1}{2}\left(r_1^2+r_2^2\right)+\omega^{-1/2}\frac{1}{r_{12}}
\label{eqn: large_omega_H}
\end{align}
This makes it clear that the high-density limit ($\omega \rightarrow \infty$) corresponds to the weakly-correlated limit, since here the Coulomb term vanishes. For large but finite $\omega$ one can treat $\omega^{-1/2}$ as a perturbation prefactor. This allows us to derive expansions for the radial wavefunction and total energy in this limit (see \cref{sec: large_omega}). 

By similar reasoning, $\omega \rightarrow 0$ corresponds to the strongly-correlated limit, and again one can use perturbation theory to derive the wavefunction and total energy\cite{Cioslowski2000}.

Given a two-electron wavefunction and its total energy, we can also extract various quantities as predicted by the exact functional. For example, the density given by the exact functional will match the analytical solution. Less trivially, we can obtain analytical solutions for the exchange-correlation potential and the Kohn-Sham energy eigenvalues (see \cref{sec: ks-functionals} for details). Moreover, we can express the Koopmans spectral correction to the exchange-correlation potential as a functional of quantities obtained by the ``base'' functional (see \cref{sec: ODD corrections} for details). 

\section{Computational Methods\label{sec: comp-methods} }
In this work we present Koopmans spectral functional calculations on Hooke's atom for $\omega=\nicefrac{1}{10}$, $\omega=\nicefrac{1}{2}$, and $\omega=10$. For the first two potentials, we compare the results against the analytic solutions. The third potential $\omega = 10$ approaches the weakly-correlated limit \cite{Matito2010}, so in lieu of an analytic solutions we rely on numerical calculations and the high-density expansions for reference results.

In order to apply Koopmans spectral functionals to Hooke's atom, we modelled Hooke's atom using the \textsc{Quantum ESPRESSO} distribution (QE) \cite{Giannozzi2009,Giannozzi2017}. The semi-local DFT calculations were performed using QE version 6.8. The ODD calculations were performed using \texttt{koopmans-kcp}, an implementation of ODD functionals built on top of QE version 4.1 \cite{Borghi2015,koopmans-kcp,Linscott2023}. 

Because QE is a DFT code that uses a plane-wave basis and assumes periodic boundary conditions, implementing the aperiodic harmonic potential of Hooke's atom within the framework of QE was not trivial. To achieve this, we implemented a potential which close to the origin is harmonic, but for asymptotic distances far from the origin is Coulombic. This is to conform with QE, which assumes that atomic potentials behave like the Coulomb potential of a point charge; that is, they should asymptotically approach $-Z_\mathrm{val}/r$, where $r$ is the distance from the nucleus and $Z_\mathrm{val}$ is the charge of the nucleus plus any pseudized electrons; in the case of hookium $Z=2$. The hookium potential is vertically shifted such that the potential vanishes for large distances, and between the harmonic and Coulombic regimes an exponential crossover makes the potential continuously differentiable.  We denote the position of the transition between the two regimes $r_c$. In all calculations the electronic density was confined well within the harmonic region of the potential.

Practically, using QE to solve Hooke's atom is excessive. For example, the system is radially symmetric and therefore one only needs to solve the one-dimensional radial problem. Additionally, using QE prevented us from exploring very large/very small values of $\omega$, to stay within the typical length- and energy-scales of real systems. Nevertheless, we opted to use QE in order for the analysis conducted here to be transferable to other more complicated systems in the future.

We perform the calculations with the LDA, PBE, PZ, KI, and KIPZ functionals. Throughout this work we will use PBE \cite{Perdew1996a} as the semi-local base functional. We performed convergence analyses to find appropriate values for the energy cutoff, the cell size, and $r_c$. Since the relevant length scales decrease with increasing omega, the convergence analysis gives omega-dependent results summarized in \cref{sec: tabulated simulation parameters}. Input files, output files and scripts for reproducing our results can be found at \href{https://doi.org/10.24435/materialscloud:mc-xr}{10.24435/materialscloud:mc-xr}.

\section{Results\label{sec: results}}
\subsection{Quasiparticle energies}

\begin{table*}
\caption{The HOMO, LUMO and LUMO+1 orbital energies as well as the total energy of Hooke's atom obtained with different functionals and different values of $\omega$. All values are in Hartree. In parentheses the difference to the reference solution is given in percent. For PBE we present results calculated via the Kohn-Sham eigenenergies (denoted $\varepsilon_i$) and via total energy differences ($\Delta$SCF). The reference solutions correspond to either analytical results (where they are available) or highly accurate numerical values from the literature \cite{Matito2010,Cioslowski2012}.} \label{tab. energies}
\def\arraystretch{1.5}
\begin{tabular}{c l c d d d d d d d d d d}
\hline
\hline
$\qquad \omega \qquad$ & \multicolumn{2}{l}{Method} &\multicolumn{2}{c}{HOMO}  & \multicolumn{2}{c}{LUMO} & \multicolumn{2}{c}{LUMO+1}& \multicolumn{2}{c}{Total}\\
\hline
\hline
    \multirow{7}{*}{$\frac{1}{10}$} & \multicolumn{2}{l}{Reference}
    & 0.3500\footnotemark[1] && 0.5594\footnotemark[3]
    && 0.5947\footnotemark[3] && 0.5000\footnotemark[5] & \\
    & \multicolumn{2}{l}{LDA}  & 0.4258 & (+21.67\%) & 0.4961 & (-11.32\%)& 0.5766 & (-3.05\%)& 0.5038 & (+0.76\%)\\
    & \multirow{2}{*}{PBE} & $\varepsilon_i$ & 0.4251 & (+21.46\%) & 0.4960 & (-11.35\%)& 0.5775 & (-2.90\%)&
    \multirow{2}{*}{0.5006\hspace{-\widthoffourzeros}} &
    \multirow{2}{*}{($+$0.12\%)\hspace{-\widthofpercent}}\\
    & & $\Delta$SCF &   0.3527 & (+0.76\%)& 0.5568 & (-0.48\%)& 0.6222 & (+4.62\%) &\\
    & \multicolumn{2}{l}{PZ}   &0.3555 & (+1.58\%)&0.4967 & (-11.21\%)& 0.5778 & (-2.86\%)& 0.5063 & (+1.26\%)\\
    & \multicolumn{2}{l}{KI}  & 0.3527 & (+0.76\%)& 0.5568 & (-0.48\%)& 0.6204 & (+4.31\%)& 0.5006 & (+0.12\%)\\
    & \multicolumn{2}{l}{KIPZ} & 0.3559 & (+1.69\%)& 0.5624 & (+0.53\%)& 0.6364 &(+7.01\%) & 0.5057  & (+0.14\%)\\
\hline
    \multirow{7}{*}{$\frac{1}{2}$} & \multicolumn{2}{l}{Reference} & 1.2500\footnotemark[1] && 2.0132\footnotemark[3] && 2.3107\footnotemark[3] && 2.0000\footnotemark[5]\\
    & \multicolumn{2}{l}{LDA}  & 1.4446 & (+15.57\%) & 1.8605 & (-7.58\%)& 2.3052 & (-0.24\%)& 2.0257 & (+1.29\%)\\
    & \multirow{2}{*}{PBE} & $\varepsilon_i$  & 1.4391 & (+15.13\%) & 1.8587 & (-7.67\%)& 2.3062 & (-0.19\%)& \multirow{2}{*}{2.0090\hspace{-\widthoffourzeros}} &
    \multirow{2}{*}{($+$0.45\%)\hspace{-\widthofpercent}}\\
    & & $\Delta$SCF &   1.2564 & (+0.52\%)& 2.0083 & (-0.24\%)& 2.4094 & (+4.27\%)&\\
    & \multicolumn{2}{l}{PZ}   &1.2563 & (+0.50\%)&1.8599 & (-7.62\%)& 2.3069 & (-0.16\%)& 2.0059 & (+0.30\%)\\
    & \multicolumn{2}{l}{KI}  & 1.2564 & (+0.52\%)& 2.0083 & (-0.24\%)& 2.4045 & (+4.06\%)& 2.0090 & (+0.45\%)\\
    & \multicolumn{2}{l}{KIPZ} & 1.2560 & (+0.48\%)& 2.0158 & (+0.13\%)& 2.4425 &(+5.71\%) & 2.0061 & (+0.31\%) \\
    \hline
    \multirow{7}{*}{10} & \multicolumn{2}{l}{Reference} & 17.4487\footnotemark[2] && 28.6898\footnotemark[4] && 37.5239\footnotemark[4] && 32.4487\footnotemark[6]\\
    & \multicolumn{2}{l}{LDA}  & 18.4501 & (+5.74\%) & 27.9979 & (-2.41\%)& 37.6990 & (+0.47\%)& 32.6962 & (+0.76\%)\\
    & \multirow{2}{*}{PBE} & $\varepsilon_i$ & 18.3814 & (+5.35\%) & 27.9621 & (-2.54\%)& 37.6717 & (+0.39\%) & \multirow{2}{*}{32.5311\hspace{-\widthoffourzeros}} &
    \multirow{2}{*}{($+$0.25\%)\hspace{-\widthofpercent}}\\
    & & $\Delta$SCF &   17.4913 & (+0.24\%)& 28.6830 & (-0.02\%)& 38.1680 & (+1.72\%)&\\
    & \multicolumn{2}{l}{PZ}   &17.4654 & (+0.10\%)&27.9654 & (-2.52\%)& 37.6746 & (+0.40\%)& 32.4504 & (+0.01\%)\\
    & \multicolumn{2}{l}{KI}  & 17.4913 & (+0.24\%)& 28.6829 & (-0.02\%)& 38.1335 & (+1.62\%)& 32.5329 & (+0.26\%)\\
    & \multicolumn{2}{l}{KIPZ} & 17.4515 & (+0.02\%)& 28.6992 & (+0.03\%)& 38.3152 &(+2.11\%) & 32.4518  & (+0.01\%)\\
    \hline
    \hline
\end{tabular}
    \footnotetext[1]{the total energy difference between the exact result for $E(N=1)$ from \cref{eqn: exact_total_one} and the exact result for $E(N=2)$ from \cref{eq: tot_tenth,eq: tot_half} (see \cref{eqn: HOMO}) }
    \footnotetext[2]{the total energy difference between the exact result for $E(N=1)$ from \cref{eqn: exact_total_one} and a  numerical result for $E(N=2)$ from Ref.~\onlinecite{Matito2010}}
    \footnotetext[3]{the total energy difference between the exact result for $E(N=2)$ from \cref{eq: tot_tenth,eq: tot_half} and  numerical results for $E(N=3)$ from Ref.~\onlinecite{Cioslowski2012} (the ground state energy when computing the LUMO; that of the first excited state for LUMO+1) }
    \footnotetext[4]{the total energy difference between  numerical results for $E(N=2)$ from Ref.~\onlinecite{Matito2010} and for $E(N=3)$ from Ref.~\onlinecite{Cioslowski2012} (the ground state energy for LUMO, that of the first excited state for LUMO+1)}
    \footnotetext[5]{exact results for $E(N=2)$ from \cref{eq: tot_tenth,eq: tot_half}}
    \footnotetext[6]{numerical result for $E(N=2)$ from Ref.~\onlinecite{Matito2010}}
\end{table*}

As discussed at the beginning of this paper, one of the main strengths of Koopmans spectral functionals is their ability to accurately predict quasi-particle energies. Thus, we will first examine the results for the HOMO, LUMO, and LUMO+1 energies; these are presented in \cref{tab. energies}. 

The semi-local functionals (LDA and PBE) dramatically overestimate the reference HOMO energy. A major improvement is achieved using the $\Delta$SCF method. This is an approach where one performs two DFT (in our case PBE) calculations and uses the resulting difference in total energy $\Delta E_i^\textrm{PBE}$ as an approximation of the quasi-particle energy. As explained before, for a small two-electron system we expect this to give a much more accurate estimate than taking the KS orbital energies. This is also what we observe in these calculations: for all three values of $\omega$, $\Delta$SCF yields relative errors for the HOMO energies that are 20 to 30 times smaller than those obtained with PBE. 

The Koopmans spectral functionals, as well as PZ, match the performance of $\Delta$SCF. This is no surprise: for KI in particular, this functional is constructed in such a way that its orbital energies match those of $\Delta$SCF. Accordingly, with KI we get an equally significant improvement over PBE. We will discuss why the KIPZ result is so similar to the KI result later when discussing the result for the exchange-correlation potential. For spin-unpolarised two-electron systems KIPZ and PZ functionals treat occupied orbitals exactly the same, provided they use the same screening parameter $\alpha$ (see \cref{sec: ODD corrections}). Since the electrons in Hooke's atom screen very little, the HOMO energy obtained with KIPZ ($\alpha \approx 0.89$, $0.94$, $0.98$ for $\omega=\nicefrac{1}{10}$, $\nicefrac{1}{2}$, and $10$ respectively) is approximately the same as the one obtained with unscreened PZ ($\alpha=1$).


The Koopmans spectral functionals are the only functionals considered in this study that provide accurate results for the triply-degenerate LUMO energy, giving relative errors that are approximately 20 times (for $\omega=\nicefrac{1}{10}$) to 80 times (for $\omega=10$) smaller than the relative errors obtained with the semilocal functionals. Also note that the PZ correction, which yielded very good results for the HOMO energy, is as inaccurate as PBE and LDA in the case of the LUMO energy. This is because --- unlike Koopmans spectral functionals --- the PZ correction only adds a correction to filled states, and not to empty ones. 

For the LUMO+1 energy, the LDA, PBE, and PZ actually outperform the Koopmans spectral functionals. This is very surprising. If we compare the PBE results (that is, both looking at the actual eigenvalue, and at the $\Delta$SCF result), the eigenvalue is closer to the reference result. This is extremely unusual, as in nearly all real systems the LDA and GGA eigenvalues tend to underestimate energy differences of empty states \cite{Parr1989a}, and the total energy differences tend to be much more reliable. Because the Koopmans orbital energies are constructed to match these differences in total energies, it is no surprise that in this case, the results obtained with the Koopmans spectral functionals are also inaccurate. As such, we believe that this result represents an outlier, rather than indicating a fundamental failure of Koopmans spectral functionals.

\subsection{The total energy and total density}
While Koopmans spectral functionals improve the description of excited states, they ought not to adversely affect ground state properties such as the total energy and total density (which are already relatively well-described by DFT). The total energies of Hooke's atom as calculated by the various functionals are listed alongside the quasiparticle energies in \cref{tab. energies}; we see that all functionals obtain total energies that are very close to the reference results. 

\begin{figure}
\includegraphics[width=\columnwidth]{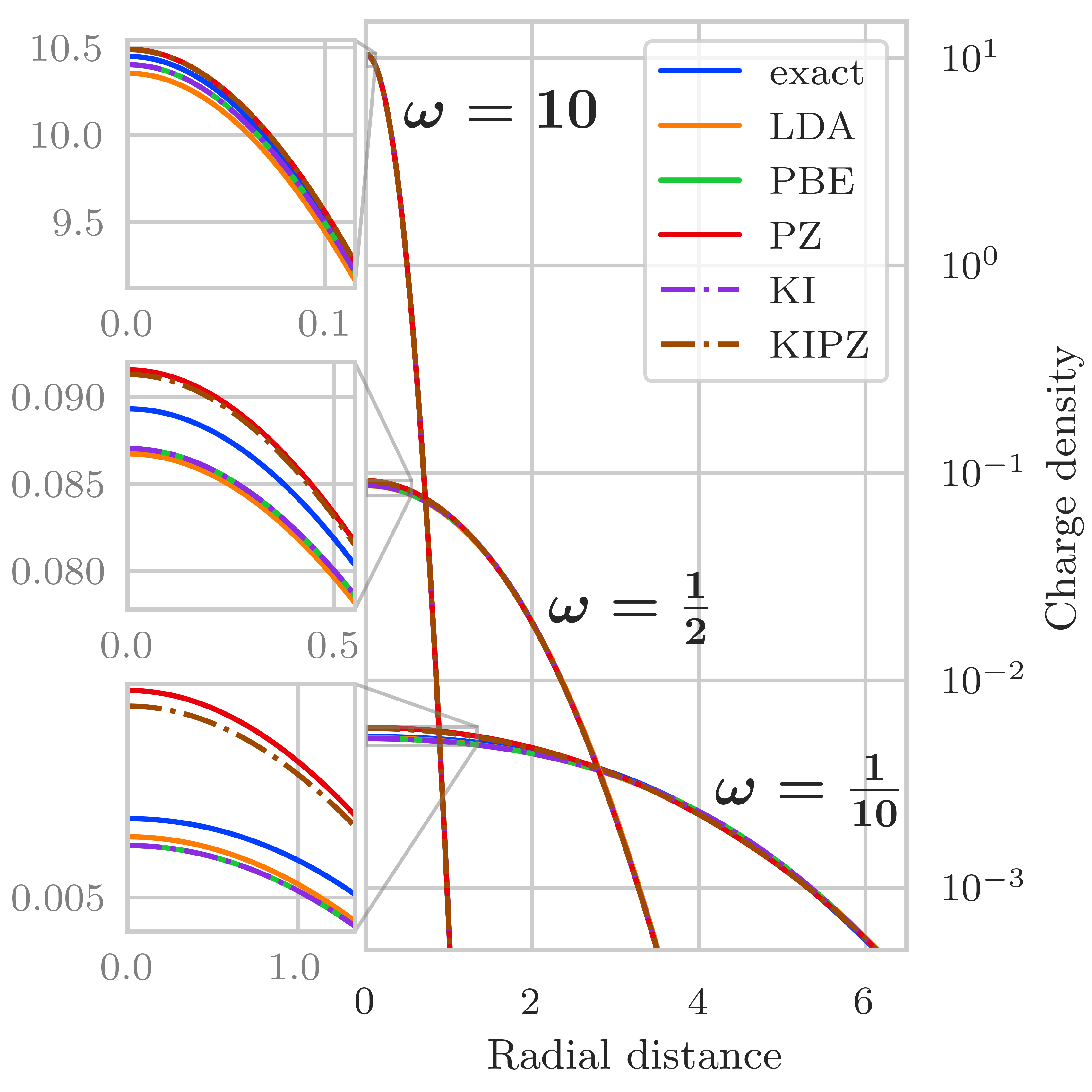}
\caption{The total electronic density as given by semi-local DFT, PZ, and Koopmans spectral functionals, as well as the exact analytical result for $\omega=\nicefrac{1}{10}$ and $\omega=\nicefrac{1}{2}$ (\cref{eqn: gs_half,eqn: gs_density_half,eqn: gs_tenth}) and the first order expansion for $\omega=10$ (\cref{eqn: large_omega_wavefunction,eqn: large_omega_f}). Atomic units are used throughout. While it may not appear to be the case, the electron density is in all cases normalized such that integrating $4\pi r^2 \rho(r)$ from zero to infinity gives two.}\label{fig: density}
\end{figure}

As for the total densities, these are presented in \cref{fig: density}. The KI density is exactly equal to the PBE density. This is expected since for integer occupations the total KI energy (and therefore also its ground state density) is equal to the total energy obtained with the underlying DFT method. By a similar reasoning the KIPZ density would be equal to the density of its underlying PZ functional if we took for both functionals the same screening parameters. However, they differ slightly, because in contrast to the KIPZ correction, the PZ correction here is unscreened. While for small distances the LDA and PBE densities are too small compared to the exact density, the PZ and the KIPZ densities are too large. On the one hand, the LDA and PBE orbitals are over-delocalized due to the self-interaction error (SIE). On the other hand, KIPZ and PZ are both methods that explicitly target the one-body SIE. In this particular case, these methods overshoot, with their densities being too localized. 

\subsection{The exchange-correlation potential}

Next we look at the exchange-correlation potential $v_{xc}(\mathbf{r})$. This is a central quantity in approximate density-functional theory since it is the only term that is unknown and must be approximated. Being one of the terms in the Kohn-Sham Hamiltonian, it directly influences the Kohn-Sham eigenvalues, so by studying the exchange-correlation potential one may understand the results for the HOMO level in more detail.

In contrast, ODDFTs do not have one such potential: each orbital is subjected to its own unique potential. Nevertheless, in the following we will compare the $v_{xc}$ potentials from DFT calculations with $v_{xc}$ plus the ODD corrective potential from ODDFT calculations. This is possible because Hooke's atom has only one doubly occupied orbital, so in this specific case (focusing for the moment on its effect on the occupied manifold) all the electrons are subject to the same local potential, the ODDFT functionals effectively become DFT functionals, and we can compare the two. We could not make such a straightforward comparison for a system with more electrons. That said, ultimately both the Kohn-Sham potential of DFT and the orbital-dependent potentials of ODDFTs are approximations of the same quantity: the electronic self-energy. The exact electronic self-energy is a non-local and dynamic quantity. The Kohn-Sham exchange-correlation potential is the best local and static approximation to the electronic self-energy~\cite{Casida1995}, while ODD corrections can be interpreted as a contribution to the dynamic but local self-energy of a discretized spectral functional theory\cite{Ferretti2014}. (Spectral functional theories approximate the self-energy with a local but dynamical quantity, with the exact spectral functional predicting the exact spectral density $\rho(\mathbf{r},\omega)$ \cite{gatti_transforming_2007}.) Thus, we can always justify comparing the Kohn-Sham exchange-correlation potential with the Kohn-Sham exchange-correlation potential plus an ODD correction by interpreting both as approximations to the exact self-energy, but for more complex systems the fact that we would then be dealing with non-local and/or dynamic quantities would make the comparison very difficult.

\begin{figure}
\includegraphics[width=\columnwidth]{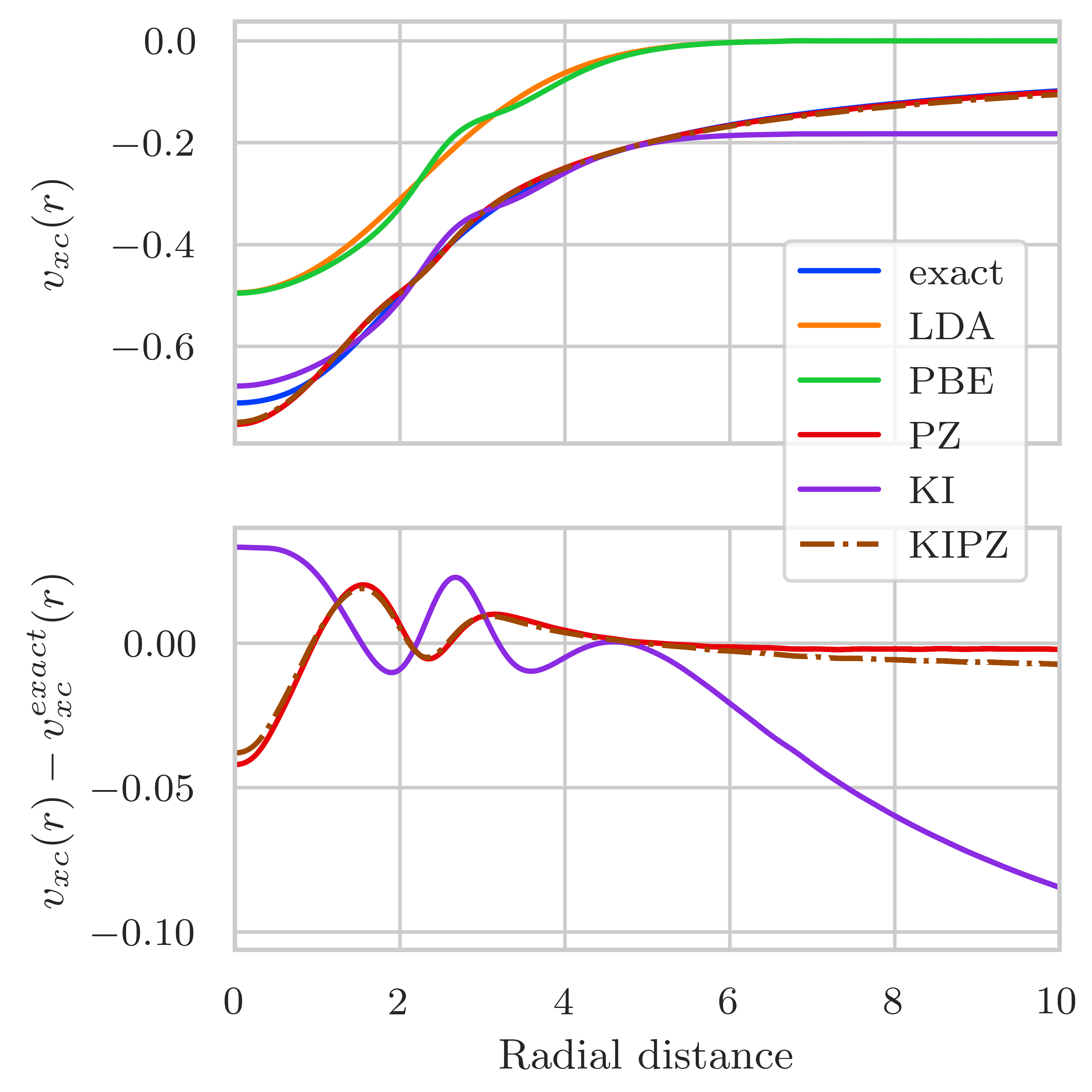}
\caption{The exchange-correlation potential of Hooke's atom, as given by semi-local DFT, PZ, and Koopmans spectral functionals. Note that for the ODDFT functionals we plot the xc potential of the base functional, plus the orbital-specific correction applied to the HOMO. We stress that this is only analogous to the corresponding DFT result because Hooke's atom is a two-electron, spin-unpolarized system.}\label{fig: xc}
\end{figure}

\Cref{fig: xc} shows the different exchange-correlation approximations in comparison to the exact exchange-correlation potential for $\omega=\nicefrac{1}{2}$ as derived in \cref{eqn: xc}.  The ODD correction for the KI, KIPZ and PZ case correspond to the formulas given in \cref{eqn: KI correction,eqn: KIPZ correction,eqn: PZ correction} respectively. The KI potential correction for filled orbitals is a constant shift to the PBE exchange-correlation potential. This is already sufficient to reproduce the exact exchange-correlation potential fairly accurately. At this point it is important to stress that all the ingredients of these Koopmans corrections are computed ab-initio. This includes the magnitude of this vertical shift that takes us from the PBE to the KI result, and it is remarkable how accurately the necessary shift is predicted by the KI functional. The largest deviation from the exact exchange-correlation potential can be seen for large radial distances. Here, the constant KI correction is not capable to correct for the incorrect asymptotic decay of the PBE exchange-correlation potential. The KIPZ potential corrects this, and also recovers the true asymptotic Coulombic behaviour for large distances. For small distances it preforms as well as the KI correction. The KIPZ and PZ results are very similar, for the same reasons that were discussed in the context of the HOMO energy.

\begin{figure}
\includegraphics[width=\columnwidth]{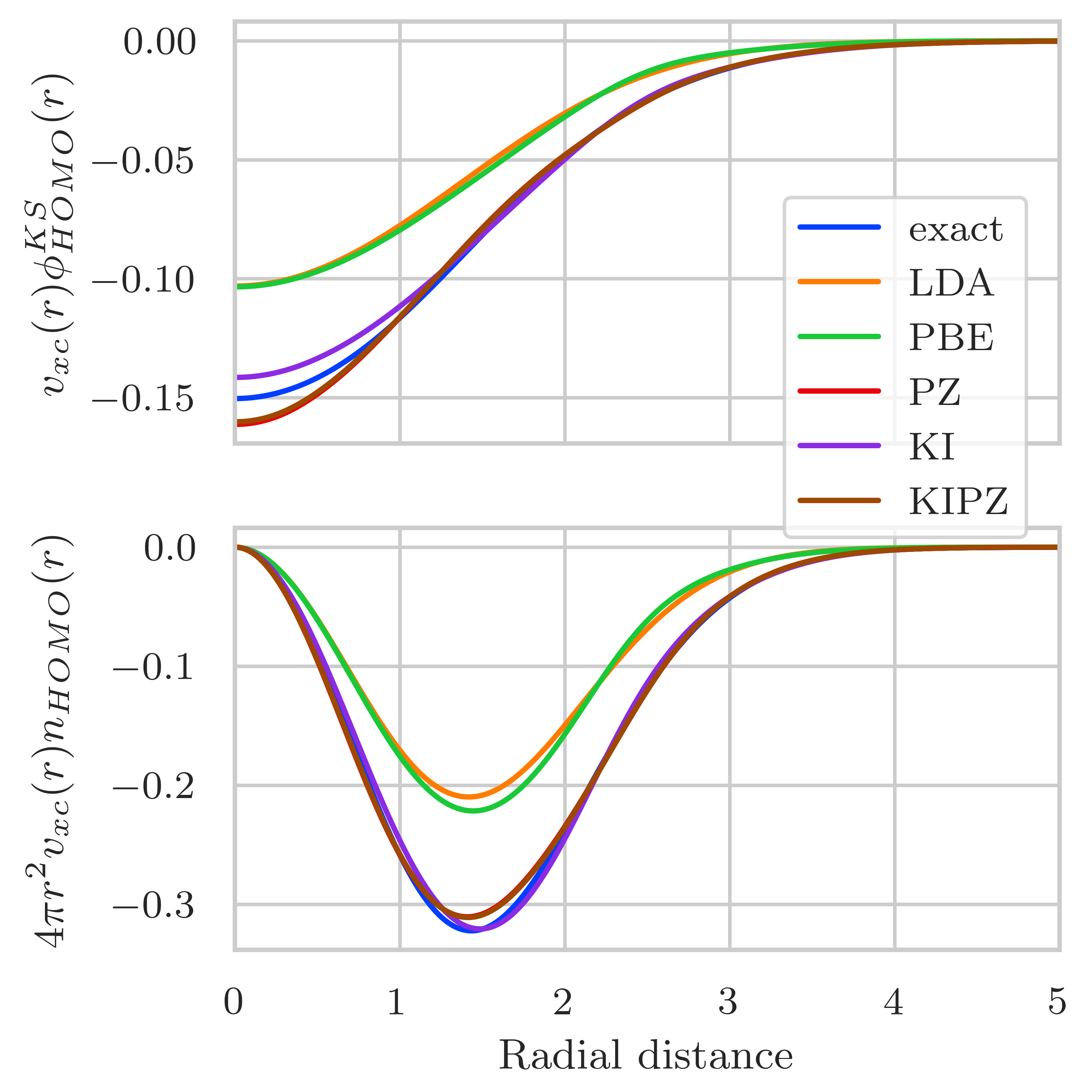}
\caption{The potential plotted in \cref{fig: xc} multiplied by the single-electron wavefunction (upper panel) and $4\pi r^2 n_{HOMO}(r)$ (lower panel).}\label{fig: both_applied}
\end{figure}

Next, we examine the exchange-correlation potential applied to the HOMO. This is presented in \cref{fig: both_applied} for $\omega=\nicefrac{1}{2}$; analgous plots for the other values of $\omega$ can be found in \cref{sec: ten_and_tenth_results}. This is a key physical quantity since during the evaluation of observables such as the total energy or the quasiparticle energies we only ever consider the potential operating upon a wavefunction, and never just the potential. For example, by looking at the Kohn-Sham equations \cref{eq: KS1} and \cref{eq: KS2}, we can directly deduce the HOMO energy by multiplying the KS equations on both sides with the ground state wavefunction and integrating over all space. To visualize the corresponding integrand, the total exchange-correlation potential (including ODD corrections for the ODDFT methods) multiplied with $4\pi r^2n_\mathrm{HOMO}(r)$ is also shown in \cref{fig: both_applied}. From this we can directly give another perspective on two results that we presented at the beginning of this section. First, the constant shift of the KI correction brings the integrand very close to the exact integrand which explains why we get a very accurate estimate of the HOMO energy with KI. Secondly, we can conclude from this figure that neither very small distances (due to the vanishing $r^2$-factor) nor very large distances (due to the vanishing electron density) contribute much to the integral. Therefore KIPZ doesn't provide a significant improvement over KI for the HOMO energy, despite the fact that KIPZ approximates the true exchange-correlation potential more accurately for large radial distances. 

The analogous exchange-correlation potentials of Hooke's atom for $\omega=\nicefrac{1}{10}$ and $\omega=10$ are plotted in \cref{fig: xc_ten_and_tenth,fig: both_applied_ten_and_tenth} in \cref{sec: ten_and_tenth_results}. Qualitatively, these plots show the same behavior as the $\omega=\nicefrac{1}{2}$ case presented in \cref{fig: xc,fig: both_applied}: the Koopmans corrections shift the DFT exchange-correlation potential downwards by the correct amount to match the exact result. The one exception to this is $\omega=10$ at large $r$, where the PZ and KIPZ potentials no longer match the reference curve (cf. the smaller values of $\omega$, where these potentials matched the reference result). It is difficult to conclude anything definitive from this, because the disagreement may simply be due to the fact that here the reference curve is only approximate (having been derived from the high-density expansion). Nevertheless, \cref{fig: both_applied_ten_and_tenth} shows that at the distances that matter most when computing the HOMO eigenvalue, the agreement remains universally excellent.

\section{Conclusion\label{sec: conclusion}}
This work shows how the Koopmans orbital energies, constructed to deliver total energy differences, provide results for the HOMO and LUMO energies that are an order of magnitude more accurate than the ones obtained with standard density-functional theory for the toy system of Hooke's atom. This accuracy matches the findings of earlier studies on real molecules and solids. We also see how these results directly relate to an excellent resemblance of the Koopmans potentials to the true exchange correlation potential. As a negative outlier, we have also seen that Koopmans spectral functionals fail to give an accurate prediction of the LUMO+1 level due to the failure of the $\Delta$SCF approach in this case. However, as mentioned before, this happens extremely rarely in real materials. 

As a final note we would like to mention that this study has some limitations due to the simplicity of Hooke's atom. For example, Koopmans spectral functionals (and orbital-density-dependent functionals more generally) give rise to two different sets of orbitals: canonical orbitals that diagonalize the Hamiltonian, and variational orbitals that minimize the total energy. For one-orbital systems such as Hooke's atom the Hamiltonian is not a matrix but a number; it follows that the canonical orbitals are trivially identical to the variational orbitals. In order to properly deconstruct and study the canonical/variational duality of ODDFTs, one would need to study a system with more electrons.

\begin{acknowledgments}
We gratefully acknowledge financial support from the Swiss National Science Foundation (SNSF -- project number 200021-179138). This research was supported by the NCCR MARVEL, a National Centre of Competence in Research, funded by the Swiss National Science Foundation (grant number 182892). We thank Andreas Adelmann, Riccardo De Gennaro, and Nicola Colonna for support and feedback. For the purpose of Open Access, a CC BY public copyright licence is applied to any Author Accepted Manuscript (AAM) version arising from this submission.
\end{acknowledgments}

\section*{Author Declarations}
\subsection*{Conflict of Interest}
The authors have no conflicts to disclose.

\section*{Data Availability}
The data that support the findings of this study are openly available in the Materials Cloud Archive, at \url{https://doi.org/10.24435/materialscloud:mc-xr}.

\appendix

\section{The high-density limit of Hooke's atom}
\label{sec: large_omega}
Solving \cref{eqn: large_omega_H} with first-order perturbation theory\cite{White1970, Cioslowski2000} for large $\omega$ yields the wavefunction (modulo a normalization constant)
\begin{align}
\Psi(r_1, r_2;\omega)= & 
\left(
1
+ (2\omega)^{-1/2} f\left(2^{-1/2} r_{12}\right) + \mathcal{O}(\omega^{-1})
\right) \nonumber \\
& \quad
\times 
e^{-(r_1^2 + r_2^2)/2}
\label{eqn: large_omega_wavefunction}
\end{align}
where
\begin{align}
f(x)= & -2 \pi^{-1 / 2}(1+\ln(2))+x^{-1} \nonumber \\
& \qquad -x^{-1} e^{x^2} \mathrm{erfc}(x)+ 2\int_0^x e^{y^2}\mathrm{erfc}(y) dy  \label{eqn: large_omega_f}
\end{align}
   
%
Note that \cref{eqn: large_omega_wavefunction} is written with the same length rescaling as \cref{eqn: large_omega_H}.

It follows that the total energy in this limit is given by
\begin{equation}
    E(\omega) = 3\omega + \left(\frac{2}{\pi}\right)^{1/2} \omega^{1/2} - \frac{2}{\pi} \left(1 - \frac{\pi}{2} + \ln{2}\right) + \mathcal{O}(\omega^{-1/2})
\end{equation}

For more details on Hooke's atom in the high-density limit, see Refs.~\onlinecite{Ivanov1999,Gill2005,Cioslowski2000} (and references therein).

\section{Potentials for Hooke's Atom}
\subsection{The exact functional \label{sec: ks-functionals}}
In the following it will be shown how exact expressions for the exchange and correlation potentials and for the HOMO energy can be obtained for Hooke's atom \cite{Kais1993}.

In the ground state of Hooke's atom, the only two occupied KS single-orbital wave functions $\phi_\mathrm{HOMO}^\mathrm{KS}$ and energies $\varepsilon^\mathrm{KS}_\mathrm{HOMO}$ are identical (one spin-up, the other spin-down). According to the theorems of Hohenberg and Kohn, the exact ground state electron density of the system is the same as the density of the non-interacting Kohn-Sham system. From this, the exact KS orbital wave function can be calculated: 
\begin{align}
\rho(\textbf{r})=2\left|\phi_\mathrm{HOMO}^\mathrm{KS}(\textbf{r})\right|^2 
\Rightarrow \phi^\mathrm{KS}_\mathrm{HOMO}(\textbf{r})=\left[\frac{1}{2}\rho(\textbf{r})\right]^{1/2}
\label{eqn: single-particle}
\end{align}
This can be employed to invert the Kohn-Sham equations
\begin{align}
&\text{KS 1: } \left[-\frac{1}{2}\nabla^2+v_\mathrm{eff}[\rho](\textbf{r})\right]\phi_j^\mathrm{KS}(\textbf{r})=\varepsilon_j^\mathrm{KS}\phi_j^\mathrm{KS}(\textbf{r}) \label{eq: KS1}\\
&\text{KS 2: } v_\mathrm{eff}[\rho](\textbf{r})=v_\mathrm{H}[\rho](\textbf{r})+v_\mathrm{ext}(\textbf{r})+v_\mathrm{xc}[\rho](\textbf{r})\label{eq: KS2}
\end{align}
and to obtain an expression for the exchange-correlation potential: 
\begin{align}
v_\mathrm{xc}[\rho](\textbf{r})=\varepsilon^\mathrm{KS}_\mathrm{HOMO}-v_\mathrm{ext}(\textbf{r})-v_\mathrm{H}[\rho](\textbf{r})-v_\mathrm{KE}[\rho]\label{eqn: xc}
\end{align}
The exact HOMO energy can be calculated as
\begin{align}
\varepsilon^\mathrm{KS}_\mathrm{HOMO}=E(N=2)-E(N=1) \label{eqn: HOMO}
\end{align}
Since 
\begin{align}
    E(N=1)=\frac{3}{2}\omega \label{eqn: exact_total_one}
\end{align}
is the energy of a single electron in a three dimensional harmonic potential, for the cases where we have an analytical solution (\cref{eq: tot_tenth,eq: tot_half}) we can obtain the exact HOMO energies:
\begin{align}
\varepsilon^\mathrm{KS}_\mathrm{HOMO}(\omega=\nicefrac{1}{10})&=\frac{7}{20}\text{ Hartree} \label{eqn: exact_HOMO_tenth} \\
\varepsilon^\mathrm{KS}_\mathrm{HOMO}(\omega=\nicefrac{1}{2}&=\frac{5}{4}\text{ Hartree} \label{eqn: exact_HOMO_half}
\end{align}
All the other terms on the right of \cref{eqn: xc} are known analytically or can be inferred by numerical integration or differentiation of the analytic expression of the ground state density:
\begin{align}
v_\mathrm{ext}(\textbf{r})&= \frac{1}{2}\omega^2r^2\\
v_\mathrm{H}&=\int d^3\textbf{r}\frac{\rho(\textbf{r}')}{|\textbf{r}-\textbf{r}'|}\\
v_\mathrm{KE}[\rho]&=-\frac{1}{2}\frac{\nabla^2\phi^\mathrm{KS}_\mathrm{HOMO}}{\phi^\mathrm{KS}_\mathrm{HOMO}}
\end{align}
and thus one can use \cref{eqn: xc} to obtain the exact exchange-correlation potential. Furthermore, for two electrons of opposite spin the exchange potential is just half of the negative of the Hartree potential
\begin{align}
v_\mathrm{x}[\rho](\textbf{r})=-\frac{1}{2}v_\mathrm{H}[\rho](\textbf{r})\label{eqn: x-pot}
\end{align}
which allows us to obtain the exact correlation potential via
\begin{align}
v_\mathrm{c}[\rho](\textbf{r})=v_\mathrm{xc}[\rho](\textbf{r})-v_\mathrm{x}[\rho](\textbf{r}) \label{eqn: c-pot}
\end{align}
\subsection{Approximate functionals}
\begin{figure}
\includegraphics[width=\columnwidth]{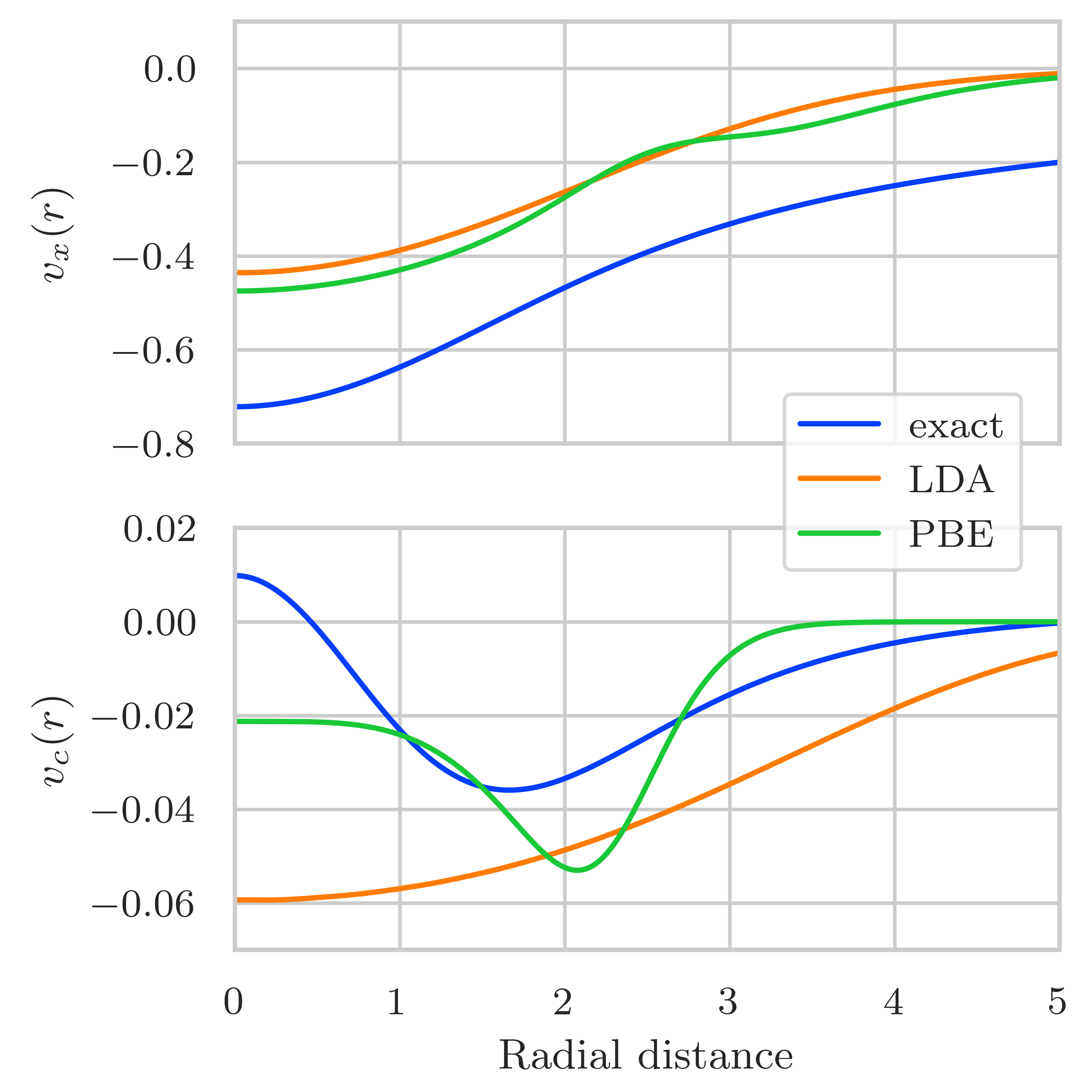}
\caption{Exchange (upper panel) and correlation potential (lower panel) obtained with LDA and PBE compared to the exact exchange potential for $\omega=\nicefrac{1}{2}$.}\label{fig: exchange-and-correlation}
\end{figure}

For the PBE and the LDA functional it is possible to extract the exchange and the correlation potential individually. These potentials can be compared to the exact expressions \cref{eqn: x-pot} and \cref{eqn: c-pot} as shown in \cref{fig: exchange-and-correlation} for $\omega=\nicefrac{1}{2}$. These plots reproduce exactly the results presented in Ref.~\onlinecite{lam_virial_1998}.

\subsection{Orbital-density-dependent corrections\label{sec: ODD corrections}}
For Koopmans spectral functionals, in addition to the exchange and correlation potentials, we must also consider the orbital-specific corrective terms. These can be obtained by taking the functional derivative with respect to the orbital density:
\begin{align}
        \hat{v}_{i\sigma}^\mathrm{Koopmans}=\frac{\delta}{\delta\rho_{i\sigma}}\sum_{j\sigma'} \Pi_{j\sigma'}^\mathrm{Koopmans}\label{eqn: KI-potential}
\end{align}
where here we have separated the spin index $\sigma$ from the orbital index $i$.

For the KI functional, the potential contribution is made up of three parts:
\begin{align}
    \frac{\delta \Pi_{i\sigma}^\mathrm{KI}}{\delta \rho_{j\sigma'}(\mathbf{r})}
    = &
    v^\mathrm{scalar}_{i\sigma}
    + \delta_{ij}\delta_{\sigma\sigma'} v^\mathrm{diag}_{j\sigma'}(\mathbf{r}) \nonumber \\
    & \quad + (1 - \delta_{ij}\delta_{\sigma\sigma'})v^\text{off-diag}_{j\sigma'}(\mathbf{r}),
\end{align}
where the scalar (i.e. $\mathbf{r}$-independent) contribution is given by
\begin{align}
    v^\mathrm{scalar}_{i\sigma}
    = & -E_\mathrm{Hxc}[\rho - \rho_i] + E_\mathrm{Hxc}[\rho - \rho_i + n_i] \nonumber \\
    & \quad - \int d\mathbf{r}' v_\mathrm{Hxc}^\sigma(\mathbf{r}',[\rho - \rho_i + n_i]) n_i(\mathbf{r'}),
\end{align}
and the diagonal, non-scalar contribution is
\begin{equation}
    v^\mathrm{diag}_{i\sigma}(\mathbf{r})
    = -v^\sigma_\mathrm{Hxc}(\mathbf{r},[\rho])
    + v^\sigma_\mathrm{Hxc}(\mathbf{r},[\rho - \rho_{i\sigma} + n_{i\sigma}]),
\end{equation}
and the off-diagonal, non-scalar contribution is
\begin{align}
    v^\text{off-diag}_{i\sigma}(\mathbf{r}) = &
    (1 - f_{i\sigma}) v^\sigma_\mathrm{Hxc}(\mathbf{r},[\rho - \rho_{i\sigma}])
    -v^\sigma_\mathrm{Hxc}(\mathbf{r},[\rho]) \nonumber \\
    & \quad + f_{i\sigma} v^\sigma_\mathrm{Hxc}(\mathbf{r},[\rho - \rho_{i\sigma} + n_{i\sigma}]).
\end{align}
Here, $f_{i\sigma}$ is the occupancy of orbital $i$ and spin $\sigma$, $n_i(\mathbf{r})$ is the density of orbital $i$, $\rho(\mathbf{r}) = f_i n_i(\mathbf{r})$ is the occupancy-dependent density, and $\rho(r) = \sum_i \rho_i(\mathbf{r})$ is the total density.

Compared to this KI correction, the KIPZ correction includes few additional terms\cite{Borghi2014}:
\begin{align}
    \frac{\delta \Pi_{i\sigma}^\mathrm{KIPZ}}{\delta \rho_{j\sigma'}(\mathbf{r})}
    = &
    \frac{\delta \Pi_{i\sigma}^\mathrm{KI}}{\delta \rho_{j\sigma'}(\mathbf{r})}
    - \bigg(
    E_\mathrm{Hxc}[n_{i\sigma}] + v^\sigma_\mathrm{Hxc}(\mathbf{r},[n_{i\sigma}]) \nonumber \\
    & - \int d\mathbf{r}' v_\mathrm{Hxc}^\sigma(\mathbf{r'},[n_{i\sigma}])n_{i\sigma}(\mathbf{r'}) \bigg) \delta_{ij} \delta_{\sigma\sigma'}
\end{align}

In the case of Hooke's atom, these corrective terms simplify dramatically. In this two-electron system there is only one filled orbital with $n_{\sigma}=\rho_{\sigma}=\nicefrac{1}{2}\rho$. Thus the general expressions above simplify to 
\begin{align}
    {v}_\sigma^\mathrm{KI}(\textbf{r})|_{f_{\sigma}=1}=&-E_\mathrm{Hxc}\left[\frac{1}{2}\rho\right]+E_\mathrm{Hxc}\left[\rho\right]\nonumber\\
    & \quad - \frac{1}{2}\int d\textbf{r}'v_\mathrm{Hxc}\left[\rho\right](\textbf{r}')\rho(\textbf{r}')\label{eqn: KI correction}\\
    {v}_\sigma^\mathrm{KIPZ}(\textbf{r})|_{f_{\sigma}=1}=&{v}_\sigma^\mathrm{KI}(\textbf{r})|_{f_{\sigma}=1}\nonumber\\
    & \quad -E_\mathrm{Hxc}\left[\frac{1}{2}\rho\right]-v_\mathrm{Hxc}\left[\frac{1}{2}\rho\right](\textbf{r})\nonumber\\
    & \qquad +\frac{1}{2}\int d\textbf{r}'v_\mathrm{Hxc}\left[\frac{1}{2}\rho\right](\textbf{r}')\rho(\textbf{r}') \nonumber\\
    & = \left(E_\mathrm{xc}\left[\rho\right]-2E_\mathrm{xc}\left[\frac{1}{2}\rho\right]\right)\nonumber\\
    & \quad +\frac{1}{2}\int d\textbf{r}' \left(v_\mathrm{xc}\left[\frac{1}{2}\rho\right]-v_\mathrm{xc}\left[\rho\right]\right)\rho(\textbf{r}') \nonumber\\
    & \qquad -v_\mathrm{Hxc}\left[\frac{1}{2}\rho\right](\textbf{r})   \label{eqn: KIPZ correction}
\end{align}
The subscript $\mathrm{Hxc}$ indicates that we take the sum of Hartree, exchange and correlation contributions. Note that the KI potential does not depend on $\textbf{r}$ in our case, i.e. it implies just a constant shift compared to the base functional. To include screening we multiply each expression above with the screening parameter $\alpha_i$ of the corresponding orbital $i$.

The PZ correction removes the $v_\mathrm{Hxc}$-potential from each orbital, i.e. the PZ orbital dependent correction is in general given by
\begin{align*}
    \hat{v}^\mathrm{PZ}_i(\textbf{r}) = -v_\mathrm{Hxc}[\rho_{i}](\textbf{r})
\end{align*}
This expression simplifies in the case of Hooke's atom: 
\begin{align}
    {v}_\sigma^\mathrm{PZ}(\textbf{r})|_{f_{\sigma}=1}&  = -v_\mathrm{Hxc}\left[\frac{1}{2}\rho\right](\textbf{r}) \label{eqn: PZ correction}
\end{align}
If we neglected orbital relaxation, i.e. if we set the screening parameters to $1$, in our case the PZ and the KIPZ potential would be identical except for a constant shift. Therefore they would yield the same density and hence the same total energy. 


\section{Simulation parameters}
\label{sec: tabulated simulation parameters}
\Cref{tab: simulation parameters} summarizes the values of key  parameters used in the \texttt{\textsc{Quantum ESPRESSO}} calculations presented in this work.
\begin{table}[t]
\caption{The simulation parameters for different values of $\omega$ obtained from convergence analyses. These parameters are discussed in \cref{sec: comp-methods}.}
\begin{ruledtabular}
    \begin{tabular}{c d d d}
         $\omega$& \multicolumn{1}{c}{energy cutoff (Ry)} & \multicolumn{1}{c}{cell size (\AA)}&  \multicolumn{1}{c}{$r_c$ (\AA)}\\
         \hline
         \hline
         $\nicefrac{1}{10}$ & 20 & 30.0 & 8.50 \\
         $\nicefrac{1}{2}$ & 50 & 12.5 & 3.50 \\
         10 & 600 & 2.6 & 0.73 \\
    \end{tabular}
    \end{ruledtabular}
    \label{tab: simulation parameters}
\end{table}

\section{Additional results}
\label{sec: ten_and_tenth_results}
\begin{figure*}[t]
\begin{subfigure}[h]{0.45\textwidth}
\includegraphics[width=\columnwidth]{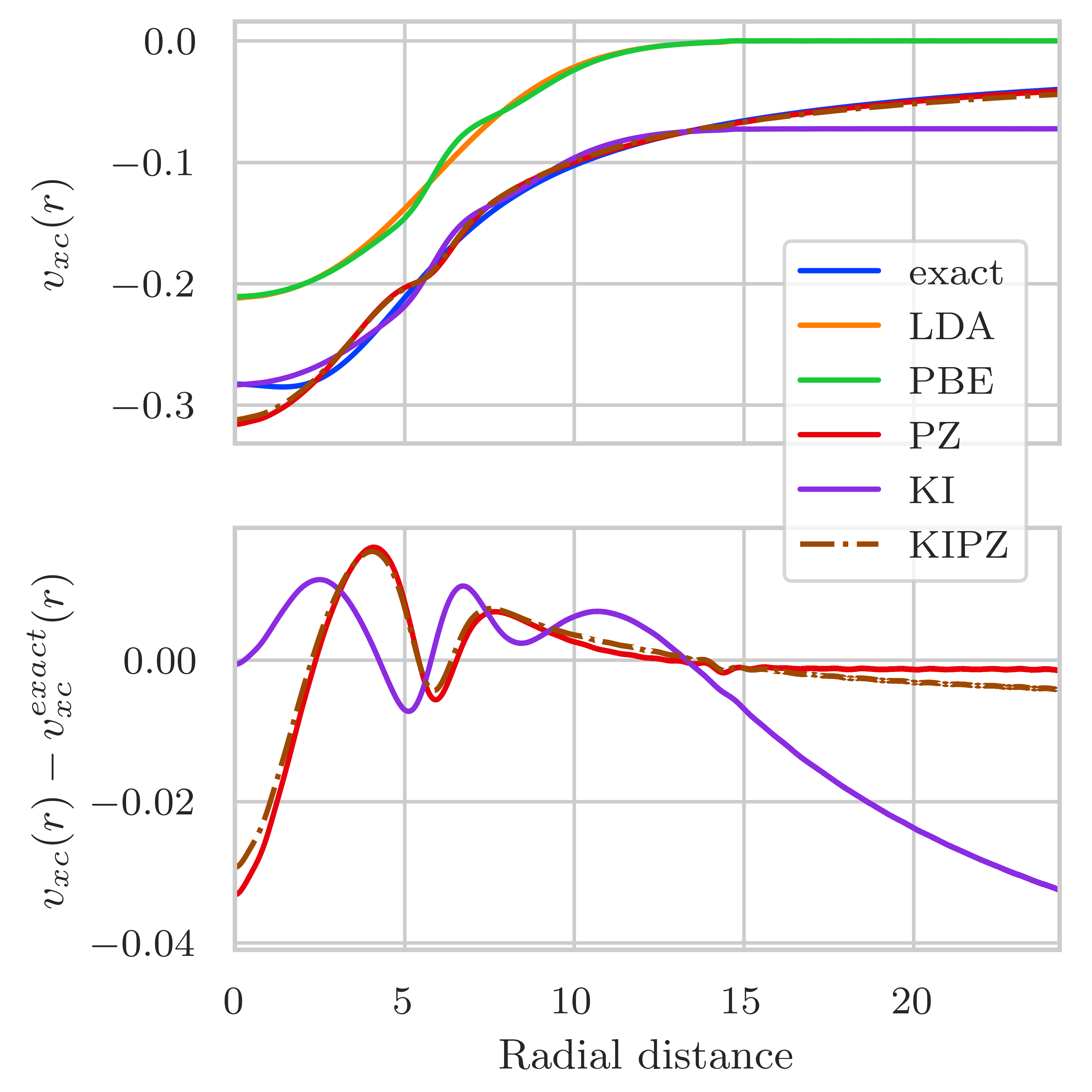}
\caption{$\omega = \frac{1}{10}$}
\label{fig: xc_tenth}
\end{subfigure}
\begin{subfigure}[h!]{0.45\textwidth}
\includegraphics[width=\textwidth]{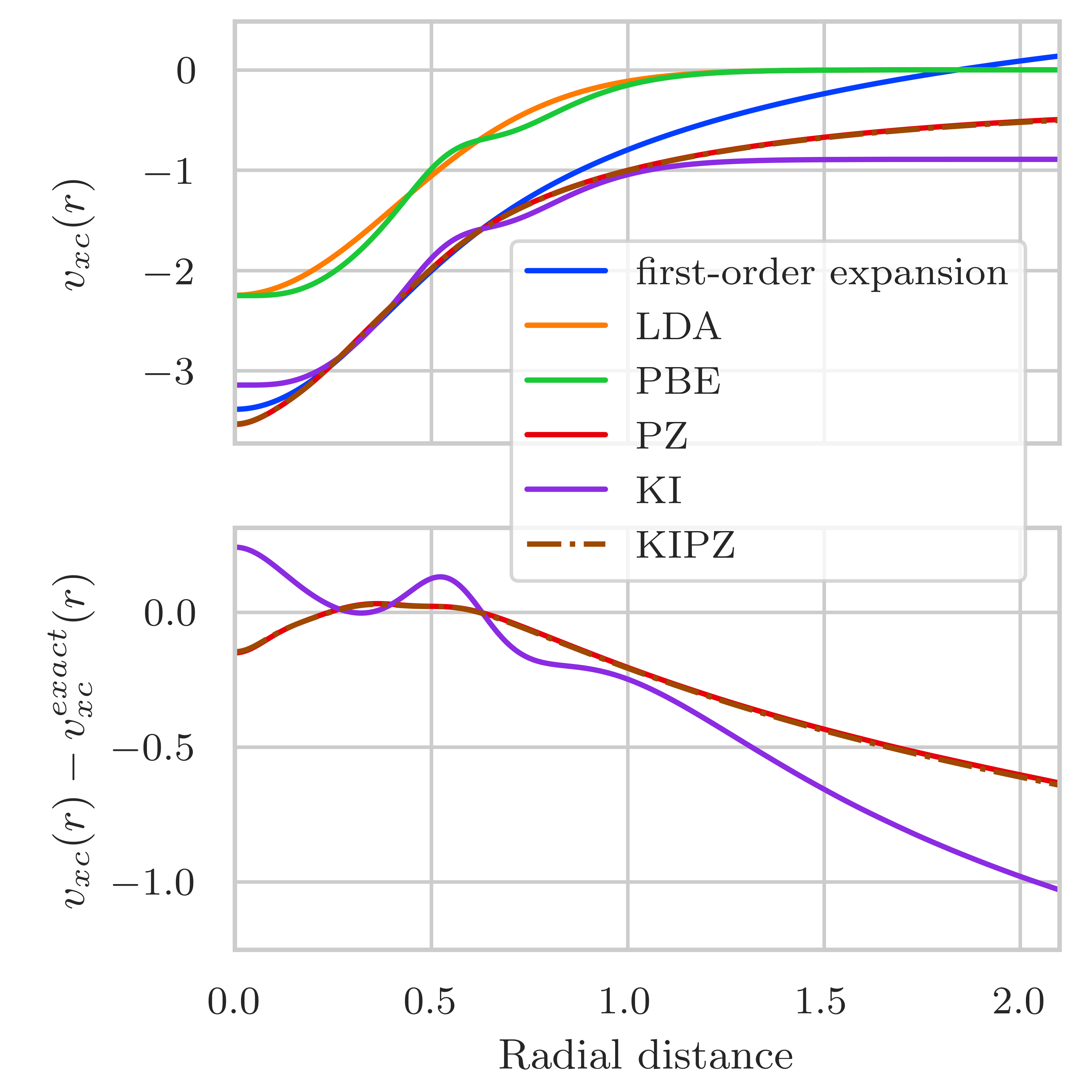}
\caption{$\omega = 10$}
\label{fig: xc_ten}
\end{subfigure}
\caption{The exchange-correlation potential of Hooke's atom with (a) $\omega=\nicefrac{1}{10}$ and (b) $\omega = 10$, as given by semi-local DFT, PZ, and Koopmans spectral functionals. For the ODDFT functionals we plot the xc potential of the base functional, plus the orbital-specific correction applied to the HOMO. For $\omega=10$ we do not have an analytical solution; instead, the reference result is derived from the high-density expansion of the wavefunction as given by perturbation theory (\cref{eqn: large_omega_wavefunction}), from which the exchange-correlation potential can be obtained as described in \cref{sec: ks-functionals}.}
\label{fig: xc_ten_and_tenth}
\end{figure*}

\begin{figure*}[h]
\begin{subfigure}[h]{0.45\textwidth}
\includegraphics[width=\columnwidth]{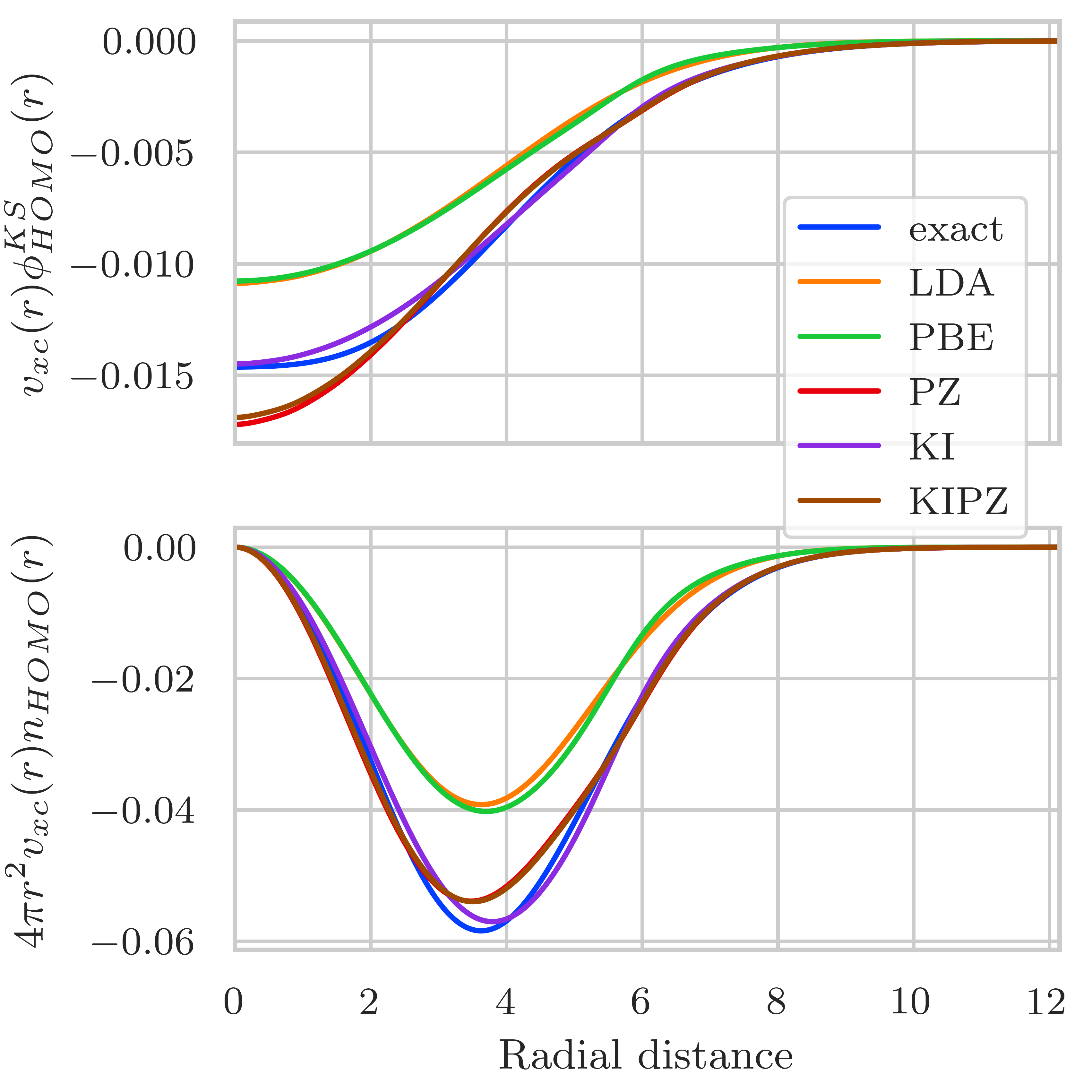}
\caption{$\omega = \frac{1}{10}$}
\label{fig: both_applied_tenth}
\end{subfigure}
\begin{subfigure}[h!]{0.45\textwidth}
\includegraphics[width=\textwidth]{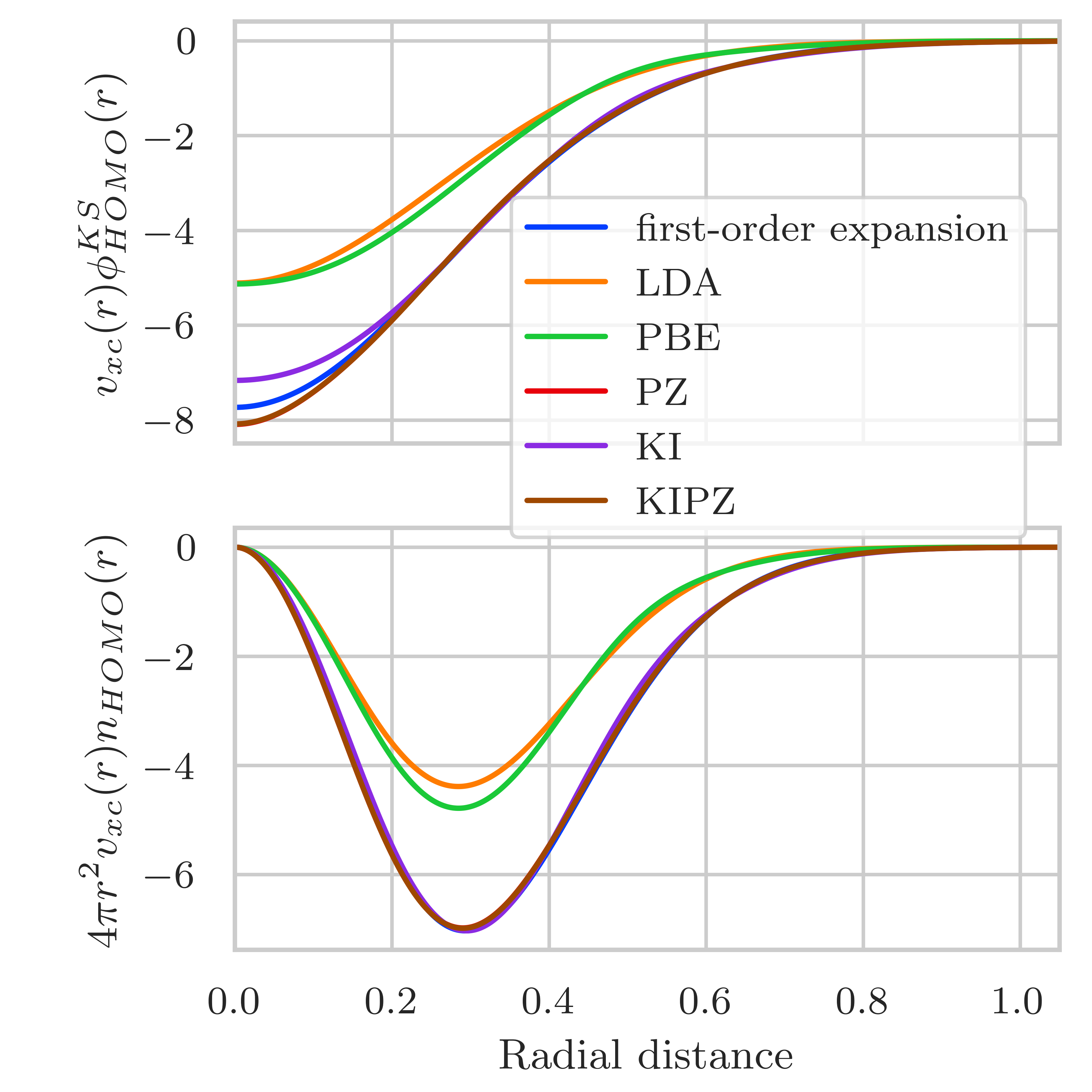}
\caption{$\omega = 10$}
\label{fig: both_applied_ten}
\end{subfigure}
\caption{The potential plotted in \cref{fig: xc_ten_and_tenth} multiplied by the single-electron wavefunction (upper panels) and $4\pi r^2 n_{HOMO}(r)$ (lower panels). As it did in \cref{fig: xc_ten_and_tenth}, the $\omega = 10$ reference curve relies on the first-order approximation in the high-density limit (\cref{eqn: large_omega_wavefunction}).}
\label{fig: both_applied_ten_and_tenth}
\end{figure*}

The exchange-correlation potentials of Hooke's atom for $\omega=\nicefrac{1}{10}$ and $\omega=10$ are plotted in \cref{fig: xc_ten_and_tenth,fig: both_applied_ten_and_tenth}. (These are analogues of \cref{fig: xc,fig: both_applied}). 

\bibliography{references.bib}
\end{document}